\documentclass[a4paper,USenglish,cleveref, autoref, thm-restate]{lites-v2021}
\pdfoutput=1 %uncomment to ensure pdflatex processing (mandatatory e.g. to submit to arXiv)
\hideLITES  %uncomment to remove references to LITES series (logo, DOI, ...), e.g. when preparing a pre-final version to be uploaded to arXiv or another public repository

\usepackage[T1]{fontenc}
\usepackage{graphicx}
\usepackage{amsmath}
\usepackage{amssymb}
\usepackage{amsfonts}
\usepackage{multirow}
\usepackage{siunitx}
\usepackage[T1]{fontenc}
\usepackage[utf8]{inputenc}
\usepackage{booktabs}
\usepackage{caption}
\usepackage{subcaption}
\usepackage[linesnumbered,lined,boxed,commentsnumbered]{algorithm2e}
\usepackage{multirow}
\usepackage{multicol}
\usepackage{prettyref}
\usepackage[inline]{enumitem}
\usepackage{subcaption}
\usepackage{booktabs, makecell, tabularx}
\newcolumntype{C}{>{\centering\arraybackslash}X} % centered version of "X" type

\usepackage{stfloats}
\usepackage{siunitx}
\usepackage{caption}
\usepackage[table]{xcolor} % http://ctan.org/pkg/xcolor

\newlist{enumnone}{enumerate*}{1}
\setlist[enumnone]{label=}
\newlist{enumalpha}{enumerate*}{1}
\setlist[enumalpha]{label=(\alph*)}

\newrefformat{tab}{Table \ref{#1}}
\usepackage{tikz}
\usetikzlibrary{shapes.geometric, arrows.meta, positioning, backgrounds}

\newcommand{\CYCLE}[1]{\eta}

\newcommand{\BSWFACTOR}{f^{BSW}}

\newcounter{challenge}

\newcounter{contribution}

\makeatletter
\renewenvironment{abstract}{%
  \vskip\bigskipamount
  \ifLineNumbers\nolinenumbers\let\linno@n\relax\fi
  \multicolsep\z@
  \columnsep\parindent
  \begin{multicols}{2}%
    [\noindent
    \rlap{\color{dagpubLineGray}\vrule\@width\textwidth\@height1\p@}%
    {\hspace*{4.5mm}\fboxsep1.5mm\colorbox{white}{\raisebox{-0.4ex}{%
    \large\selectfont\sffamily\bfseries\abstractname}}}%
    \vskip3\p@]%
  \fontsize{9}{12}\selectfont
  \noindent\ignorespaces}
  {%
  \end{multicols}%
  \ifx\linno@n\relax\linenumbers\fi
  \protected@write\@auxout{}{\string\gdef\string\@pageNumberEndAbstract{\thepage}}%
}
\def\ps@authorversionfirstpage{%
  \def\@evenhead{\large\sffamily\bfseries
                 \llap{\hbox to0.5\oddsidemargin{  \ifx\@hideDagpub\@undefined\ifx\@ArticleNo\@empty\textcolor{red}{XX}\else\@ArticleNo\fi:\fi\thepage\hss}}\hfil}%
  \def\@oddhead{\large\sffamily\bfseries\hfil
                \rlap{\hbox to0.5\oddsidemargin{\hss  \ifx\@hideDagpub\@undefined\ifx\@ArticleNo\@empty\textcolor{red}{XX}\else\@ArticleNo\fi:\fi\thepage}}}%
  \let\@oddfoot\@empty
  \let\@evenfoot\@empty
  \let\@mkboth\markboth
}
\makeatother

\title{Shedding Light onto Safety Integrity Level and Basic Software Constraints in a Real-World Automotive Application: Case Study with Driverator Framework} %TODO Please add

\titlerunning{Case Study with Driverator Framework} %TODO optional, please use if title is longer than one line

\author{Tobias Denzinger}{CARIAD SE, Ingolstadt, Germany}{tobias.denzinger@audi.de}{}{}
\author{Matthias Becker}{KTH Royal Institute of Technology Stockholm, Sweden}{mabecker@kth.se}{https://orcid.org/0000-0002-1276-3609}{}
\author{Peter Ulbrich}{Technische Universität Dortmund, Germany}{peter.ulbrich@tu-dortmund.de}{https://orcid.org/0000-0002-4224-9205}{}

\authorrunning{T. Denzinger, M. Becker, P. Ulbrich} %TODO mandatory. First: Use abbreviated first/middle names. Second (only in severe cases): Use first author plus 'et al.'

\Copyright{Tobias Denzinger, Matthias Becker, Peter Ulbrich}

\ccsdesc[100]{\textcolor{red}{Replace ccsdesc macro with valid one}} %TODO mandatory: Please choose ACM 2012 classifications from https://dl.acm.org/ccs/ccs_flat.cfm 

\keywords{Automotive software, case study, ASIL, AUTOSAR BSW} %TODO mandatory; please add comma-separated list of keywords

\nolinenumbers %uncomment to disable line numbering

\Volume{1}
\Issue{1}
\Article{42}
\DateSubmission{Date of submission}
\DateAcceptance{Date of acceptance}
\DatePublished{Date of publishing}
\SectionAreaEditor{LITES section area editor}
\begin{document}

\maketitle
\thispagestyle{authorversionfirstpage}

\begin{abstract}
Automotive electronic control units (ECUs) constitute highly complex embedded systems comprising hundreds of functions, numerous software components, and multiple mutually dependent tasks. A common architectural pattern in such systems is represented by cause–effect chains. While existing research has extensively addressed the temporal analysis and optimization of these chains—particularly with respect to data age and reaction time—other non-functional properties have received comparatively limited attention.
Among these properties, the safety integrity level (SIL) plays a central role in system design, as it directly constrains task allocation and colocation decisions. Inappropriate sharing of functions or the interleaving of tasks with different safety classifications may jeopardize the integrity of safety-critical functionality. Moreover, AUTOSAR basic software (BSW), including the operating system, runtime environment, communication services, and diagnostic components, introduces additional design complexity that depends on task characteristics and SIL classifications. Memory requirements further intensify this challenge, since heterogeneous memory architectures and SIL-dependent constraints impose strict limitations on feasible task mappings.
This case study provides a detailed characterization of a real-world automotive application, with particular emphasis on SIL constraints, AUTOSAR BSW impact, and memory requirements.
We introduce the Driverator Framework, which enables researchers to generate application instances that reflect the characteristics of realistic industrial systems. Unlike existing case studies, Driverator incorporates additional system-relevant aspects, including SIL classifications, basic software overheads, and memory consumption.

\end{abstract}

\begin{mdframed}[
  backgroundcolor=dagpubLightGray,
  linewidth=0,
  innertopmargin=6pt,
  innerbottommargin=6pt,
  innerleftmargin=8pt,
  innerrightmargin=8pt,
  skipabove=0.8\baselineskip,
  skipbelow=1.0\baselineskip
]
\small
\noindent\textbf{Tool availability.}
The Driverator tool and accompanying case-study material are archived as
\cite{Driverator} at
\href{https://doi.org/10.17877/TUDODATA-2026-MOR12ARE}{\nolinkurl{https://doi.org/10.17877/TUDODATA-2026-MOR12ARE}}.
\end{mdframed}

% -----------------------------------------------
\section{Introduction} \label{sec:introduction}
% -----------------------------------------------

In recent years, automotive electronic control units (ECUs) have undergone a substantial increase in complexity, driven by the integration of multiple applications, advanced driver-assistance functions, and the stringent architectural requirements imposed by functional safety standards such as ISO~26262~\cite{ISO26262}. Modern automotive ECUs typically integrate tens to hundreds of interconnected software components (SW-Cs), whose design is constrained by non-functional requirements, including timing, memory consumption, and, in particular, Safety Integrity Levels (SILs). SIL classifications have a direct impact on architectural design decisions, as they impose strict constraints on function colocation, task allocation, and memory isolation. These constraints become especially relevant in heterogeneous hardware architectures, where safety properties may differ across processing cores and memory regions. An inappropriate synthesis or allocation of tasks with different SIL classifications can compromise system integrity, resulting in safety risks as well as architectural inefficiencies.

Automotive systems are further characterized by their reliance on standardized software platforms such as AUTOSAR~\cite{AUTOSARRTE}. AUTOSAR provides basic software (BSW) services, including operating systems, runtime environments (RTEs), communication stacks, and diagnostic services. While these components constitute essential infrastructure for automotive applications, they also contribute significantly to overall system complexity and resource demand. The overhead introduced by BSW depends on functional and non-functional properties, thereby creating a complex interdependence between software architecture, safety requirements, and system performance.

A common architectural pattern in automotive software is the cause-effect chain, which organizes tasks into data-flow sequences from sensor inputs to actuator outputs. These chains represent safety- and performance-relevant functional paths in which data is propagated across tasks, potentially operating at different execution rates~\cite{kramer2015real}. A key design criterion for such chains is their end-to-end timing behavior, which must satisfy predefined constraints to ensure the required quality of service. In particular, age constraints define the maximum admissible time between the acquisition of an input value by the first task of a chain and the production of the corresponding output value by the final task~\cite{Feiertag2008ACF}. Incorporating SIL constraints into the design of chains further increases architectural complexity, as both application SW-Cs and BSW components must be allocated to hardware resources while respecting safety, timing, and memory-related requirements.

% Figure 1 ---------------------------------------
\begin{figure}[tb]
    \centering
    \includegraphics[width=0.67\columnwidth]{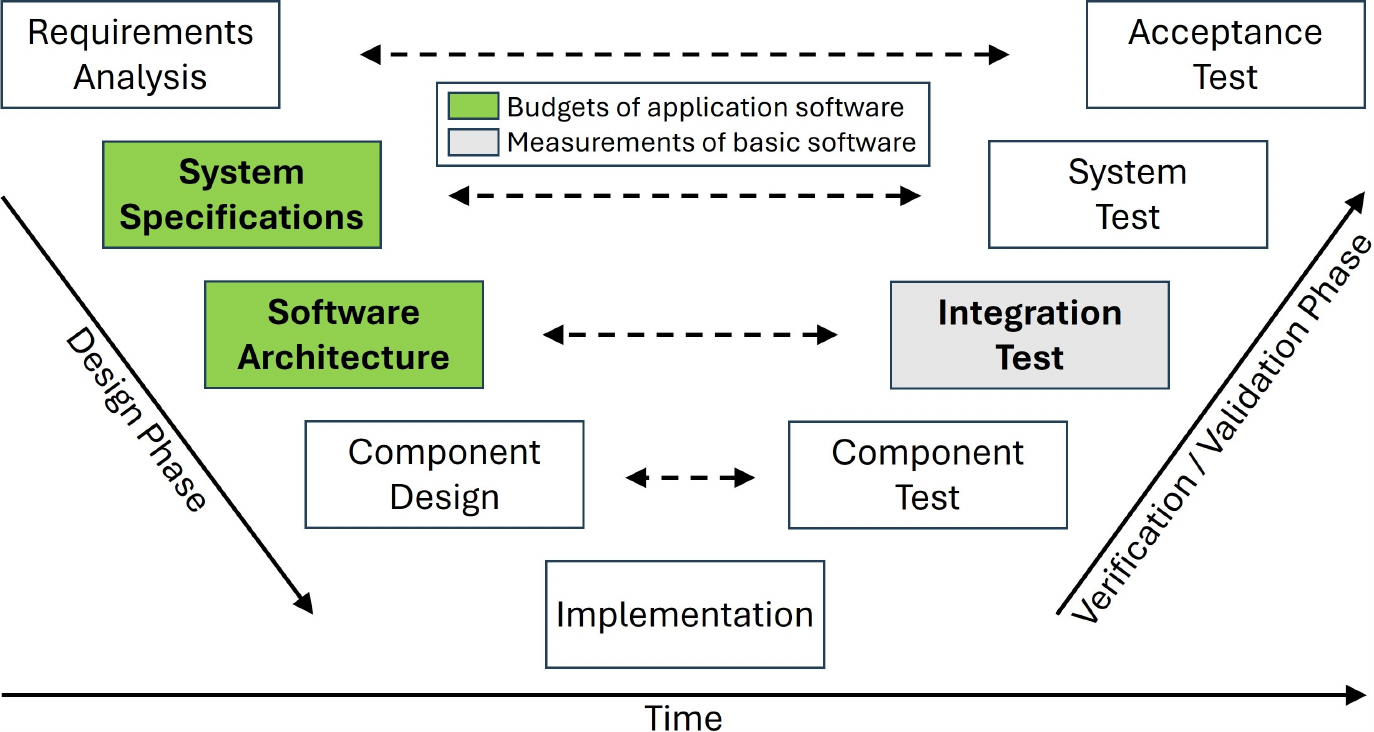}
    \caption{Extraction of respective application and basic software characteristics according to the highlighted V-model phases.}
    \label{fig:vModel}
\end{figure}
% -----------------------------------------------

In contrast to prior work such as~\cite{kramer2015real}, this paper provide more detailed description of the applications including SIL classifications, BSW overheads, and memory impact in automotive systems. Our objective is to identify correlations between these design aspects, with a particular focus on the early design phase. At this stage, the functional scope and system specifications are typically known, while individual development teams still work on separate functions, SW-Cs, and BSW components without a finalized integration into a concrete hardware platform. 
With reference to the V-model illustrated in Fig.~\ref{fig:vModel}, the proposed work initially focuses on the left-hand side of the development process. At this stage, early architectural decisions exert a substantial influence on the subsequent system design, while only conservative estimates of timing and resource budgets are typically available.
As the process progresses towards the right-hand side of the V-model, the software design becomes increasingly refined based on empirical measurements. This refinement aims to achieve tight bounds between the allocated budgets and the corresponding worst-case execution times (WCETs).

Although substantial research has addressed automotive systems and cause-effect chains, including their characterization, timing analysis, and scheduling, and despite extensive work on mixed-criticality systems, such as the survey by Burns and Davis~\cite{Burns}, an important gap remains. Existing case studies rarely provide a holistic characterization in which safety levels, basic software, and memory requirements are treated as first-class design dimensions. To address this gap, Section~\ref{sec:casestudy} presents a case study of a real-world automotive motion and drive system that explicitly quantifies SIL constraints, BSW contributions, and memory requirements. This characterization addresses our first challenge by providing detailed statistical data and thereby establishes a foundation for generating realistic artificial test systems for evaluation. The resulting insights are intended to support both the evaluation conducted in this paper and future studies on safety-aware automotive software architecture.

% -----------------------------------------------
\section{Automotive Applications} \label{sec:AutomotiveApplications}
% -----------------------------------------------

According to the AUTOSAR automotive software framework~\cite{AUTOSARRTE}, the software architecture of an electronic control unit (ECU) is organized into a layered platform comprising the application layer, the Runtime Environment (RTE), the Basic Software (BSW) layer, and the underlying microcontroller hardware. The application layer consists of software components (SW-Cs), which encapsulate application-level functionality and provide a modular abstraction of automotive functions. In the context of functional safety, atomic software components are typically associated with an Automotive Safety Integrity Level (ASIL), as defined by ISO~26262~\cite{ISO26262}. This classification captures the criticality of the respective functionality and imposes corresponding requirements on software design, integration, and execution. Each SW-C contains at least one cyclic runnable, which represent the smallest schedulable units of application behavior. Runnables may be invoked during system initialization, periodically according to predefined activation patterns, or in response to events. In particular, cyclic runnables are commonly released periodically in accordance with AUTOSAR timing specifications~\cite{AUTOSARTiming}, thereby forming the basis for schedulable systems.

Communication between runnables is realized through standardized AUTOSAR interaction mechanisms, mainly sender-receiver communication and client-server calls. These interactions are mediated by the RTE, which abstracts application software from the underlying BSW. By decoupling SW-Cs from concrete communication mechanisms and hardware-specific details, the RTE supports modular and portable component-based software development. It therefore plays a central role in coordinating data exchange between application components and basic software services. However, this abstraction also introduces execution overhead and additional resource demands. These overheads depend on the communication pattern, the involved SW-Cs, the activation rates of runnables, and the associated ASIL classifications. Hence, the RTE must be considered not only as a functional middleware layer but also as a relevant contributor to timing, memory, and safety-related system properties.

The AUTOSAR BSW layer provides standardized infrastructure services required for ECU operation and is subdivided into service components, ECU abstraction, microcontroller abstraction, and complex drivers. These modules implement fundamental functionality such as the operating system (OS), communication stacks, diagnostic services, memory services, I/O access, and hardware-dependent drivers. Each BSW module may comprise multiple system-related runnables that interact with application-level runnables and contribute to the overall workload of the ECU. In typical AUTOSAR configurations, runnables with identical or compatible activation patterns are mapped to OS tasks, although multiple tasks with the same activation rate may coexist within a system to satisfy separation, priority, or safety requirements. The resulting tasks are scheduled by the AUTOSAR OS using fixed-priority scheduling, as specified in~\cite{AUTOSAROSSpec}. Tasks generally execute until completion and may only be preempted by tasks with higher priority. They are activated either time-synchronously at predefined rates or sporadically by specific events, such as communication interrupts or diagnostic requests. As a result, the mapping of application and BSW runnables to tasks constitutes a key architectural decision that directly affects timing behavior, memory consumption, and the preservation of ASIL-related separation constraints.

% -----------------------------------------------
\section{Approach from Benchmark Abstraction to Driverator Framework} \label{sec:BenchmarkGeneration}
% -----------------------------------------------

The benchmark considered in this work is derived from a set of industrial requirements for a real-world automotive application and is deliberately abstracted to comply with intellectual property (IP) restrictions. As a consequence, detailed descriptions of the implemented functionalities, exact software-budget values, and corresponding measurement data cannot be disclosed. To overcome these limitations while still enabling a meaningful scientific evaluation, this paper presents a representative case study in which the relevant system specifications are statistically synthesized.

% Figure 2 ---------------------------------------
\begin{figure}[h!]
    \centering
    \resizebox{0.9\columnwidth}{!}{\begin{tikzpicture}[x=1pt,y=1pt,line cap=round,line join=round]
  \useasboundingbox (0,0) rectangle (612,186.236);

  \definecolor{linegray}{RGB}{87,94,99}
  \tikzset{
    figureline/.style={draw=linegray,line width=1.0pt,fill=white},
    cylinderfill/.style={fill=white,draw=none},
    cylinderoutline/.style={draw=linegray,line width=1.0pt,fill=none},
    arrowline/.style={draw=black,line width=1.2pt,fill=white},
    figtext/.style={font=\sffamily\fontsize{11.4}{13.7}\selectfont,align=center,text=black},
    cylindertext/.style={font=\sffamily\fontsize{11.4}{13.7}\selectfont,align=center,text=black},
    labeltext/.style={font=\sffamily\fontsize{12.0}{14.4}\selectfont,align=center,text=black}
  }

  % Critical IP: stacked database symbol.
  \path[cylinderfill] (19,43) rectangle (105,96);
  \path[cylinderfill] (62,43) ellipse [x radius=43, y radius=10];
  \path[cylinderfill] (62,96) ellipse [x radius=43, y radius=10];
  \draw[cylinderoutline] (19,43) -- (19,96);
  \draw[cylinderoutline] (105,43) -- (105,96);
  \draw[cylinderoutline] (62,43) ellipse [x radius=43, y radius=10];
  \draw[cylinderoutline] (62,96) ellipse [x radius=43, y radius=10];

  \path[cylinderfill] (10,49) rectangle (96,102);
  \path[cylinderfill] (53,49) ellipse [x radius=43, y radius=10];
  \path[cylinderfill] (53,102) ellipse [x radius=43, y radius=10];
  \draw[cylinderoutline] (10,49) -- (10,102);
  \draw[cylinderoutline] (96,49) -- (96,102);
  \draw[cylinderoutline] (53,49) ellipse [x radius=43, y radius=10];
  \draw[cylinderoutline] (53,102) ellipse [x radius=43, y radius=10];

  \path[cylinderfill] (0,55) rectangle (86,108);
  \path[cylinderfill] (43,55) ellipse [x radius=43, y radius=10];
  \path[cylinderfill] (43,108) ellipse [x radius=43, y radius=10];
  \draw[cylinderoutline] (0,55) -- (0,108);
  \draw[cylinderoutline] (86,55) -- (86,108);
  \draw[cylinderoutline] (43,55) ellipse [x radius=43, y radius=10];
  \draw[cylinderoutline] (43,108) ellipse [x radius=43, y radius=10];
  \node[cylindertext] at (43,82) {RealTime\\Application};

  % Abstraction method arrow.
  \draw[arrowline] (128,71) -- (192,71) -- (192,54) -- (226,88) -- (192,122) -- (192,105) -- (128,105) -- cycle;
  \node[figtext] at (166,88) {Abstraction\\Method};

  % IP boundary.
  \draw[black,line width=1.2pt,dash pattern=on 6pt off 4pt] (245,34) -- (245,145);

  % Curved arrow to correlation coefficients.
  \draw[black,line width=1.2pt,-{Stealth[length=6pt,width=6pt]}]
    (166,105) .. controls (168,137) and (207,143) .. (253,138);

  % Correlation coefficients box.
  \draw[black,line width=0.8pt,dash pattern=on 2.5pt off 2.5pt,fill=white] (253,119) rectangle (402,159);
  \node[labeltext] at (327.5,139) {Subject-specific\\correlation coefficients};

  % Downward arrow and system properties box.
  \draw[black,line width=1.2pt,-{Stealth[length=6pt,width=6pt]}] (327.5,119) -- (327.5,91);
  \draw[figureline] (273,51) rectangle (382,94);
  \node[figtext] at (327.5,72.5) {System\\Properties via\\Case Study};

  % Deployment arrow.
  \draw[arrowline] (405,71) -- (469,71) -- (469,54) -- (503,88) -- (469,122) -- (469,105) -- (405,105) -- cycle;
  \node[figtext] at (441,88) {Deployment};

  % Driverator database symbol.
  \path[cylinderfill] (517,52) rectangle (603,104);
  \path[cylinderfill] (560,52) ellipse [x radius=43, y radius=10];
  \path[cylinderfill] (560,104) ellipse [x radius=43, y radius=10];
  \draw[cylinderoutline] (517,52) -- (517,104);
  \draw[cylinderoutline] (603,52) -- (603,104);
  \draw[cylinderoutline] (560,52) ellipse [x radius=43, y radius=10];
  \draw[cylinderoutline] (560,104) ellipse [x radius=43, y radius=10];
  \node[cylindertext] at (560,78) {Driverator\\Framework};

  % IP labels.
  \node[labeltext] at (111,22) {Critical IP};
  \node[labeltext] at (442,22) {Uncritical IP};
\end{tikzpicture}}
    \caption{Overview of methodical abstraction based on a real-world application without IP constraints.}
    \label{fig:approach}
\end{figure}
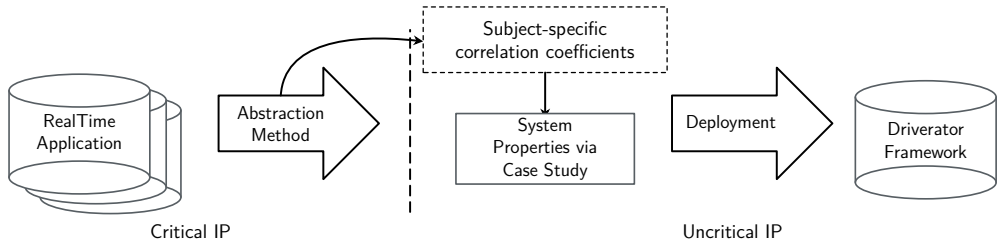
% -----------------------------------------------

Figure~\ref{fig:approach} illustrates the proposed methodology for transferring insights from a real-world industrial application, subject to IP restrictions, to the publicly available case study described in \prettyref{sec:casestudy}. Since the original data and detailed functional descriptions cannot be disclosed, the abstraction process is based on statistically aggregated system characteristics rather than on proprietary implementation details. 
Specifically, subject-related properties of the industrial benchmark are characterized using correlation coefficients. To model the underlying distributions of these characteristics, Weibull distributions are employed due to their flexibility in representing non-uniform, skewed, and heterogeneous data patterns that are commonly observed in automotive software systems. The fidelity of the resulting abstraction is evaluated by comparing the generated case-study characteristics against the original industrial benchmark using correlation-based similarity metrics. In addition, we developed a coherent system configuration tool, referred to as Driverator, which enables the construction of application structures for the case study. The archived Driverator tool, together with the accompanying case-study material, is made publicly available via DOI~\cite{Driverator}. Furthermore, the Driverator framework is designed to be scalable with respect to the underlying hardware and software platform, thereby supporting its use for timing analysis, design-space exploration, and system-level optimization.

In the early development phase, the primary objective is to assess the feasibility of the system design, as discussed in \prettyref{sec:introduction}. At this stage, the application software is designed in the absence of concrete measurement data, requiring the use of conservative timing budgets under a tight-bound assumption. In addition, the available specifications of the basic software (BSW) are typically preliminary and provide only limited timing information, which is refined progressively during later implementation phases. To address this uncertainty, we propose an estimation concept that quantifies the expected timing impact of BSW based on observations from previous development cycles. This approach exploits the fact that a substantial portion of BSW behavior remains largely consistent across projects with comparable implemented features. Consequently, the relevant application-level indicators are characterized by the interaction among SIL, basic software (BSW), and memory-related contributions. These dimensions collectively define the basis for the proposed case study and are discussed in detail in the following section.

% -----------------------------------------------
\section{Case Study: SIL, BSW, and Memory in a Real-World Automotive Vehicle Driving System} \label{sec:casestudy}
% -----------------------------------------------

This section presents the application characteristics of a real-world motion and drive controller as a representative core component of an automotive system. The analysis relies on conservative system-budget estimates available in the early development phase, prior to the availability of detailed implementation-specific measurement data.

% -----------------------------------------------
\subsection{Properties of Application Software}
\label{sec:AppSwProperties}
% -----------------------------------------------

The AUTOSAR application layer represents the uppermost layer of the AUTOSAR software architecture and contains the application software that realizes vehicle-level functionality. It consists of software components (SW-Cs), each implementing specific functional behavior and interacting with other components through standardized interfaces. The number and granularity of SW-Cs are primarily determined by the functional architecture and by the allocation of functions to the underlying execution platform. Thus, the SW-C composition reflects both the technical decomposition of vehicle functions and the constraints imposed by the target platform.

% TABLE 1 ---------------------------------------
\begin{table}[h!]
\caption{Attributes of Software Components}\label{tab:SWC-mem}
\centering
%\footnotesize
\setlength\tabcolsep{2.0em}
\begin{tabular}{ c c c c c c }
    \toprule
    \bfseries{ASIL} &
    \bfseries{Share} &
    \multicolumn{2}{c}{\textbf{ROM (kB)}} &
    \multicolumn{2}{c}{\textbf{RAM (kB)}} \\
    \cmidrule(rl){3-4}\cmidrule(rl){5-6} &&
    Min. & Max. & Min. & Max. \\
    \midrule
    QM & 41 \% & 6 & 255 & 0.5 & 29 \\
    A & 7 \% & 37 & 140 & 1 & 22 \\
    B & 14 \% & 15 & 228 & 0.5 & 17 \\
    C & 10 \% & 18 & 168 & 1 & 26 \\
    D & 28 \% & 21 & 252 & 1 & 23 \\
    \bottomrule
\end{tabular}
\end{table}
% -----------------------------------------------

Table~\ref{tab:SWC-mem} summarizes the main characteristics of the software components (SW-Cs) responsible for implementing application-specific functionality. For each SW-C, the table reports the distribution of the assigned Automotive Safety Integrity Levels (ASILs), together with the corresponding memory requirements expressed as ROM and RAM intervals.
The ASIL classification ranges from ASIL~D, which denotes the highest level of automotive risk and consequently requires the most rigorous assurance measures, to Quality Management (QM), which applies to functions without identified automotive safety hazards and therefore without safety requirements governed by the ISO~26262 safety lifecycle. The intermediate ASIL levels represent progressively lower degrees of hazard severity and corresponding assurance effort.

The analyzed SW-C set is dominated by QM-classified components, which account for 41\% of the total distribution. ASIL~D components constitute the second-largest group, representing 28\%, thereby indicating that a substantial fraction of the application-specific functionality is safety-relevant and subject to stringent development constraints. Across the considered safety classifications, the memory requirements span from 6~kB to 255~kB for ROM and from 0.5~kB to 29~kB for RAM. These ranges provide a structured characterization of the resource demands associated with each safety class and enable an assessment of their impact on the overall system architecture.
When compared with the baseline feature set, the considered SW-C attributes yield a correlation factor of 0.5174. This result indicates a moderate positive association, suggesting that the selected SW-C characteristics are meaningfully related to the baseline configuration, although they do not fully explain its variation.

% TABLE 2 ---------------------------------------
\begin{table}[h!]
\caption{Characteristics of Runnables in Application Layer}\label{tab:run-est}
\centering
%\footnotesize
\setlength\tabcolsep{1.5em}
\begin{tabular}{ c c c c l }
    \toprule
    \bfseries Period (ms) & \multicolumn{2}{c}{\textbf{WCET ($\mu$s)}} & \bfseries{Share} & \bfseries{ASIL} \\
    \cmidrule(rl){2-3} & Min. & Max. \\
    \midrule
    1 & 15 & 290 & 5 \% & QM, B, D \\ 
    5 & 10 & 725 & 31 \% & QM, A, B, C, D \\
    10 & 22 & 1,218 & 25 \% & QM, B, C, D \\
    20 & 56 & 1,733 & 12 \% & QM, B, C, D \\
    50 & 125 & 1,902 & 8 \% & QM, A, B, D \\
    100 & 228 & 2,447 & 10 \% & QM, C, D \\
    200 & 191 & 3,988 & 2 \% & QM, A \\
    500 & 403 & 6,176 & 4 \% & QM, D \\
    1000 & 1,209 & 9,200 & 3 \% & QM \\
    \bottomrule
\end{tabular}
\end{table}
% -----------------------------------------------

~\prettyref{tab:run-est} summarizes the runtime-specific budget characteristics of the considered SW-Cs and reports a correlation coefficient of 0.4322 with respect to the baseline feature set. The listed parameters include activation periods, worst-case execution-time (WCET) intervals, and the corresponding safety classifications. The distribution ratios indicate the share of runnables assigned to the application layer, with the dominant activation periods occurring at 5 ms (31\%) and 10 ms (25\%). Further notable contributions are observed at 20 ms (12\%) and 100 ms (10\%).

In most cases, each software component (SW-C) typically comprises an initialization runnable for parameter configuration and a cyclic runnable for periodic functional execution. Additional cyclic runnables are introduced only when required by specific functional features, timing constraints, or distinct execution contexts. Runnable activation is predominantly periodic and follows predefined execution rates, whereas event-triggered or interrupt-related activations are limited to selected execution paths requiring asynchronous or condition-dependent execution.
The runtime environment schedules the runnables according to predefined task mappings and activation periods.
Because runnables are structurally bound to their parent SW-C, this relationship is maintained during architectural decomposition and safety analysis. Accordingly, the SW-C provides the scope for assigning and propagating safety-related attributes (ASIL) to its associated runnables, thereby ensuring consistency between the software architecture, timing model, and safety classification.

% -----------------------------------------------
\subsection{Communication via Cause-Effect Chains}
\label{sec:chainProperties}
% -----------------------------------------------

To account for the orchestration of the considered communication paradigms, cause-effect chains are explicitly incorporated into the analysis. These critical chains comprise application-specific runnables that may exhibit either identical or heterogeneous activation patterns and may be distributed across multiple tasks. An activation pattern defines the number of runnable releases associated with each runnable within a given chain and therefore characterizes the temporal structure by which data dependencies propagate through the system.

% TABLE 3 ---------------------------------------
\begin{table}[h!]
\caption{Configuration of Cause-Effect Chains} \label{tab:act-pattern}
\centering
%\footnotesize
\setlength\tabcolsep{1.7em}
\begin{tabular}{ c c c c } 
\toprule
\multicolumn{2}{c}{\textbf{Cause-Effect Chain}} &
\multicolumn{2}{c}{\textbf{Activation Pattern}} \\
\cmidrule(rl){1-2}\cmidrule(rl){3-4} 
Activation Pattern & Share & Set of Runnables & Share \\
\midrule
\multirow{2.0}{0em} {1} & \multirow{2.0}{3em} {\centering 65 \%} & 2 & 40 \% \\
\multirow{3.25}{0em} {2} & \multirow{3.25}{3em} {\centering 25 \%} & 3 & 25 \% \\
\multirow{4.5}{0em} {3} & \multirow{4.5}{3em} {\centering 10 \%} & 4 & 20 \% \\
 & & 5 & 10 \% \\
 & & 6 & 5 \% \\
\bottomrule
\end{tabular}
\end{table}
% -----------------------------------------------

In summary, each cause-effect chain comprises between two and 18 runnables and may include one to three distinct activation patterns. The number of critical chains is determined by the implemented application features, safety-critical constraints and the characteristics of the underlying platform. Moreover, different chains may share common runnables, resulting in overlapping functional and timing dependencies. The configuration reported in \prettyref{tab:act-pattern} summarizes the relevant attributes of the considered cause-effect chains.
Within each cause-effect chain, runnables and their corresponding tasks are arranged according to a predefined execution order to ensure compliance with the specified end-to-end latency constraints. The resulting data-age bounds are derived using chain-specific hyperperiods, with minimum and maximum scaling factors of 1.8 and 4.9, respectively. Tasks are predominantly activated time-synchronously according to the periods of their assigned runnables. However, selected tasks may also be triggered by special events and executed under either preemptive or cooperative scheduling semantics.

In the context of cause-effect chains, interactions among cyclic runnables are modeled through inter-runnable communication mechanisms. As illustrated in \prettyref{tab:communication}, the communication direction is represented by assigning sender runnables to the rows and receiver runnables to the columns. The corresponding arrows indicate whether data exchange occurs between adjacent runnable periods. The highest communication volumes are typically observed between runnables with identical activation periods and are highlighted in the table. Communication is implemented using dedicated read and write labels associated with the respective runnables, following the abstraction introduced in \cite{kramer2015real}. Depending on the task mapping, these data dependencies result in either intra-task or inter-task communication.
In the proposed abstraction, runnable interactions are therefore represented through read/write label-based communication across the corresponding task context. The involved runnables may operate at identical or heterogeneous sampling rates, depending on timing requirements, data dependencies, and real-time constraints. This representation enables a systematic characterization of communication dependencies between cyclic runnables and supports the analysis of task-level interactions induced by the underlying runnable-to-task mapping.

% TABLE 4 --------------------------------------------
\begin{table}[h!]
    \begin{center}
      \caption{Matrix of Inter-Runnable Communication} \label{tab:communication}

    \begin{tabular}{ >{\centering\arraybackslash}m{1.1cm} | >{\centering\arraybackslash}m{0.7cm} | >{\centering\arraybackslash}m{0.7cm} | >{\centering\arraybackslash}m{0.7cm} | >{\centering\arraybackslash}m{0.7cm} | >{\centering\arraybackslash}m{0.7cm} | >{\centering\arraybackslash}m{0.7cm} | >{\centering\arraybackslash}m{0.7cm} | >{\centering\arraybackslash}m{0.7cm} | >{\centering\arraybackslash}m{0.7cm} }
    \hline
    {\bf Period (ms)} & {1} & {5} & {10} & {20} & {50} & {100} & {200} & {500} & {1000} \\[0ex] 
    \hline
    {1} & \cellcolor{gray!15} \LARGE \rotatebox[origin=c]{270}{$\Lsh$} & \LARGE \rotatebox[origin=c]{270}{$\Lsh$} & \LARGE \rotatebox[origin=c]{270}{$\Lsh$} &  & & & & & \\[0ex] 
    \hline
    {5} & \LARGE \rotatebox[origin=c]{270}{$\Lsh$} & \cellcolor{gray!15} \LARGE \rotatebox[origin=c]{270}{$\Lsh$} & \LARGE \rotatebox[origin=c]{270}{$\Lsh$} & \LARGE \rotatebox[origin=c]{270}{$\Lsh$} & \LARGE \rotatebox[origin=c]{270}{$\Lsh$} & & & & \\[0ex] 
    \hline
    {10} & \LARGE \rotatebox[origin=c]{270}{$\Lsh$} & \LARGE \rotatebox[origin=c]{270}{$\Lsh$} & \cellcolor{gray!15} \LARGE \rotatebox[origin=c]{270}{$\Lsh$} &\LARGE \rotatebox[origin=c]{270}{$\Lsh$} & \LARGE \rotatebox[origin=c]{270}{$\Lsh$} & & & & \\[0ex] 
    \hline
    {20} & & \LARGE \rotatebox[origin=c]{270}{$\Lsh$} & \LARGE \rotatebox[origin=c]{270}{$\Lsh$} & \cellcolor{gray!15} \LARGE \rotatebox[origin=c]{270}{$\Lsh$} & & \LARGE \rotatebox[origin=c]{270}{$\Lsh$} & &  & \\[0ex] 
    \hline
    {50} & & & \LARGE \rotatebox[origin=c]{270}{$\Lsh$} & \LARGE \rotatebox[origin=c]{270}{$\Lsh$} & \cellcolor{gray!15} \LARGE \rotatebox[origin=c]{270}{$\Lsh$} & \LARGE \rotatebox[origin=c]{270}{$\Lsh$} & & \LARGE \rotatebox[origin=c]{270}{$\Lsh$} & \\[0ex] 
    \hline
    {100} & & & & & \LARGE \rotatebox[origin=c]{270}{$\Lsh$} & \cellcolor{gray!15} \LARGE \rotatebox[origin=c]{270}{$\Lsh$} & \LARGE \rotatebox[origin=c]{270}{$\Lsh$} & \LARGE \rotatebox[origin=c]{270}{$\Lsh$} & \\[0ex] 
    \hline
    {200} & & & & & & \LARGE \rotatebox[origin=c]{270}{$\Lsh$} & \cellcolor{gray!15} \LARGE \rotatebox[origin=c]{270}{$\Lsh$} & & \LARGE \rotatebox[origin=c]{270}{$\Lsh$} \\[0ex] 
    \hline
    {500} & & & & & & & & \cellcolor{gray!15} \LARGE \rotatebox[origin=c]{270}{$\Lsh$} & \LARGE \rotatebox[origin=c]{270}{$\Lsh$}\\[0ex] 
    \hline
    {1000} & & & & & & \LARGE \rotatebox[origin=c]{270}{$\Lsh$} & & \LARGE \rotatebox[origin=c]{270}{$\Lsh$} & \cellcolor{gray!15} \LARGE \rotatebox[origin=c]{270}{$\Lsh$}\\[0ex] 
    \hline
    
  \end{tabular}
  \label{tabular:UKJPNdata}
  \end{center}
\end{table}
%---------------------------------------------

% -----------------------------------------------
\subsection{Impact of Basic Software}
\label{sec:bswProperties}
% -----------------------------------------------

The continuously increasing number of implemented features intensifies the interaction between application software and AUTOSAR Basic Software (BSW), including the RTE, communication stacks, OS, and diagnostics, thereby increasing runtime costs and integration complexity. 
As discussed in \cite{Vector2016}, the concurrent integration of safety-related and non-safety-related software further increases runtime overhead and architectural complexity. 
These effects can be mitigated through appropriate BSW partitioning measures, which separate software of different criticality levels and reduce interference between safety-relevant and non-safety-relevant execution paths.

% TABLE 5 ---------------------------------------
\begin{table}[h!]
\caption{Task Properties of Basic Software Extensions} \label{tab:bswExtensions}
\centering
%\footnotesize
\setlength\tabcolsep{2.25em}
\begin{tabular}{ c c c c }
    \toprule
     \bfseries{Count of Runnables} & \multicolumn{2}{c}{\textbf{WCET Ratio of BSW Extensions}} \\
     \cmidrule(r l){2-3}
     & Min. & Max. \\
    \midrule
    1 & 24 \% & 57 \% &\\
    2 & 19 \% & 48 \% &\\
    3 & 15 \% & 39 \% &\\
    4 & 12 \% & 30 \% &\\
    5 & 10 \% & 22 \% &\\
    $\geq$ 6 & 8 \% & 14 \% &\\
    \bottomrule
\end{tabular}
\end{table}
% ------------------------------------------------

In general, the application-dependent basic software (BSW) contribution, e.g., communication services, can be modeled either as task-local execution overhead or as separate BSW tasks. Table~\ref{tab:bswExtensions} provides a statistical approximation of task-level properties considering the impact of BSW extensions, which depends on the task composition and the number of feature-specific runnables assigned to each task. 
The abstracted BSW ratio decreases with task size, ranging from 24\%--57\% for single-runnable frames to 8\%--14\% for frames with at least six runnables.
The considered BSW-related task properties exhibit a correlation factor of 0.5042. 
Increasing the number of cyclic runnables concatenated within a task reduces the relative amount of BSW interaction overhead, since shared service invocations and scheduling-related effects can be amortized over a larger accumulated runnable workload. 
This behavior is captured by the BSW extension ratios, which quantify the additional WCET contribution of BSW services relative to the accumulated runnable WCET under consistent ASIL constraints. These extension costs must be incorporated into the extended task execution times, as they directly affect the overall processor utilization.

Essential basic software (BSW) functionality, including diagnostic services, operating-system services, and hardware-dependent drivers, is executed by dedicated BSW tasks. These tasks are assigned to discrete safety levels and are differentiated according to their activation pattern, namely periodic or sporadic activation patterns. \prettyref{tab:bswTask-wcet} summarizes the relevant characteristics of the considered BSW tasks. 
The BSW task periods span from 1~ms to 200~ms, with 5~ms and 10~ms tasks forming the dominant shares of 31\% and 34\%, respectively. The observed relationship is described by a correlation coefficient of 0.4189.
Each BSW task comprises system-related runnables executed in a predefined sequence, ensuring deterministic behavior and correct interaction of safety-relevant operations within the basic software layer.
The interaction between the BSW task set and the BSW extension set reflects a complementary design trade-off: efficiently integrating BSW functionality as application-local extensions reduces the need for separate BSW tasks, e.g., for communication services. 
Thus, the resulting BSW task configuration is determined by the allocation of BSW functionality between task-local extensions and dedicated system-level BSW tasks, thereby directly influencing the execution-time demand of the underlying system.

% TABLE 6 ---------------------------------------
\begin{table}[h!]
\caption{Attributes of Basic Software Task Budgets}\label{tab:bswTask-wcet}
\centering
%\footnotesize
\setlength\tabcolsep{1.3em}
\begin{tabular}{ c c c c l }
    \toprule
    \bfseries Period (ms) & \multicolumn{2}{c}{\textbf{WCET ($\mu$s)}} & \bfseries{Share} & \bfseries{ASIL} \\
    \cmidrule(rl){2-3} & Min. & Max. \\
    \midrule
    1 & 26 & 168 & 7 \% & QM, D \\ 
    2 & 18 & 385 & 6 \% & QM, D \\
    5 & 21 & 1,014 & 31 \% & QM, A, B, C, D \\
    10 & 38 & 1,430 & 34 \% & QM, A, B, C, D\\
    20 & 37 & 1,698 & 6 \% & QM, B, D \\
    50 & 41 & 972 & 8 \% & QM, D \\
    100 & 95 & 588 & 3 \% & QM \\
    200 & 14 & 114 & 5 \% & QM \\
    \bottomrule
\end{tabular}
\end{table}
% -----------------------------------------------

Each basic software (BSW) task contains a set of system-related runnables, ranging from a few to up to 200, which share a common activation pattern and execute according to a predefined order. Determinism shall be ensured by preserving this order across all task activations.
In summary, the BSW configuration includes approximately 6,000 system-level runnables distributed across a modern multi-core microcontroller.
This scale reflects significant architectural evolution compared with engine control units reported about one decade ago, which comprised up to 1,500 runnables \cite{kramer2015real}. Moreover, the aggregated BSW memory requirements, ranging statistically from 634 kB to 1,258 kB per lockstep-specific core, impose a non-negligible demand on the hardware resources.

% -----------------------------------------------
\section{Driverator System Configuration} \label{sec:generator}
% -----------------------------------------------

The Driverator \cite{Driverator} is controlled by a system generator that instantiates software properties according to the case-study characteristics and target platform, with reproducibility ensured through deterministic seed assignment. Initially, each software component (SW-C) and cyclic runnable is assigned a predefined safety integrity level (SIL); the mapping is iterated until each SW-C contains at least one SIL-compatible cyclic runnable and all runnables are assigned to SW-Cs. To model conservative application-software budgets, SW-C memory demands and pessimistic runnable execution-time budgets are statistically sampled between the median and upper quartile of the underlying distributions, while runnables are annotated with SIL-compliant period indices. Basic Software (BSW) contributions are modeled analogously using pessimistic budgets; depending on the applied task-merging strategy, BSW extensions are added to the WCET of affected tasks, whereas aggregated BSW tasks are characterized by statistically selected period indices and execution-time values from the same percentile range.

As discussed in Section~\ref{sec:introduction}, software estimates become increasingly accurate along the right-hand side of the V-model; consequently, pessimistic early-stage budgets are refined into tighter runtime and memory estimates by sampling between the median and standard-deviation range of the original pessimistic distributions.
Cause-effect chains are generated from predefined templates specifying activation patterns and runnable counts, with cyclic runnables permitted to participate in multiple chains to represent overlapping timing dependencies. Each chain must satisfy its end-to-end latency constraint. The allocation target is the AURIX$^{TM}$ TC39x family, where the analysis is restricted to the four lockstep cores with dedicated RAM for high-SIL workloads; the two performance cores are excluded. Interference is not modeled explicitly but is assumed to be mitigated by configuration or covered by conservative WCET assumptions.

% -----------------------------------------------
\section{Challenges} \label{sec:challenges}
% -----------------------------------------------

Automotive software development is particularly challenging in early design stages, where non-functional requirements are typically available only as abstract specifications rather than measurement-based values. Consequently, runtime and memory demands are often represented by conservative budgets that approximate feasible upper bounds until tighter estimates become available in later development phases. This uncertainty is further amplified by the basic software, whose detailed design and measurement data are usually not yet available, making its runtime and memory impact difficult to assess.
Compliance with functional-safety standards, such as ISO~26262~\cite{ISO26262}, imposes additional constraints on the mapping of runnables to tasks. In particular, freedom from interference in time and space must be ensured when integrating software functions on a shared multi-core platform. Task configurations therefore have to be designed such that runnable-level timing requirements and end-to-end constraints of cause-effect chains are satisfied, provided that the corresponding tasks remain schedulable.
As a result, early-stage architectural decisions must jointly address timing predictability, safety isolation, BSW overhead, memory consumption, and end-to-end latency constraints on multi-core systems. Balancing these partially conflicting objectives under limited design information constitutes the architect's dilemma.

The main \textbf{requirements} that need to be met are described below:
\begin{description}
	\item[R1] Each runnable is assigned to exactly one periodic task with implicit deadlines; a task may contain multiple runnables. Tasks are mapped under partitioned FP scheduling.
	\item[R2] Clustering/merging must preserve SIL separation and chain context and must respect data-age constraints.
	\item[R3] All runnables of the same SW-C must be co-located on one core; per-core ROM/RAM capacities must not be exceeded.
	\item[R4] BSW overheads are modeled as task-local extensions (via $\BSWFACTOR(\cdot)$) and as global BSW tasks with per-core memory envelopes.
\end{description}

% -----------------------------------------------
\section{Conclusion and Future Work} \label{sec:conclusion}
% -----------------------------------------------

This paper investigates the characteristics of complex automotive applications during early design phases.
We introduce an abstraction method that transfers an IP-critical benchmark into the Driverator framework, thereby enabling scalable analyses across diverse system configurations. The resulting system characteristics are derived from a real-world automotive drive-system case study. Beyond timing-related metrics, the study further considers SIL constraints, AUTOSAR basic software overheads, and memory requirements.
In this way, the Driverator framework provides a ready-to-use approach for generating realistic system models of varying sizes that represent applications in early development phases on the left side of the V-model.

Furthermore, the automotive software development process involves additional challenges in the context of real-time systems, offering relevant opportunities for future research and analysis.

\begin{enumerate}
  \item Establishing tight bounds between conservative software budgets used in early design phases and measurement-based realistic worst-case execution time (WCET) estimates obtained in later development stages.
  \item Considering the internal design of BSW-specific cause–effect chains, ranging from a small number to a few hundred system-related runnables executed in a predefined order, deterministic behavior shall be ensured while satisfying end-to-end latency constraints.
  \item Performing a precise analysis of heterogeneous hardware platforms that differ in lockstep and non-lockstep mechanisms, and assessing their impact on SIL-constrained cause-effect chains in multi-core systems.
\end{enumerate}

%%
%% Bibliography
%%

%% Please use bibtex, 

%\bibliography{lites-v2021-sample-article}
%\bibliographystyle{splncs04}
\bibliography{References.bib}

\appendix

\end{document}